\documentclass[11pt,twoside]{article}
\usepackage{vdbbook}

\resetcounters

\def\Lo {\ifmmode{{\rm ~L}_\odot}\else{~L$_\odot$}\fi}
\def\Mo {\ifmmode{{\rm ~M}_\odot}\else{~M$_\odot$}\fi}

\begin{document}

\allowtitlefootnote

\title{The brightest ULIRG: watching the birth of a quasar}
\author{
Ray P. Norris$^1$, Minnie Y. Mao$^{1,2,3}$, Emil Lenc$^{1,4}$, Bjorn Emonts$^1$, Rob G. Sharp$^5$
\affil{$^1$ CSIRO Astronomy \& Space Science, PO Box 76, Epping, NSW 1710, Australia\\
$^2$National Radio Astronomy Observatory, Socorro, NM 87801, USA\\
$^3$University of Tasmania, Hobart, 7001, Australia \\
$^4$The University of Sydney, NSW 2006, Australia\\
$^5$Research School of Astronomy \& Astrophysics, The Australian National University, Weston Creek, ACT 2611, Australia\\
}}

\begin{abstract}
The extreme ULIRG F00183-7111 has recently been found to have a radio-loud AGN with jets in its centre, representing an extreme example of the class of radio-loud AGNs buried within dusty star-forming galaxies. This source appears to be a rare example of a ULIRG glimpsed in the (presumably) brief period as it changes from ``quasar mode'' to ``radio mode'' activity. Such transition stages probably account for many of the high-redshift  radio-galaxies and extreme high-redshift ULIRGs, and so this object at the relatively low redshift of 0.328 offers a rare opportunity to study this class of objects in detail. We have also detected the CO signal from this galaxy with the ATCA, and here describe the implications of this detection for future ULIRG studies.
\end{abstract}

\section{Introduction}
\label{intro}

In the early 1980's, the IRAS satellite discovered a number of unexpectedly bright far-infrared (FIR) sources. A number of groups around the world \citep{aaronson84, houck85, allen85} quickly found that they were at high redshifts, and thus of enormous total luminosity, rivalling and even exceeding that of quasars. \citet{harwit} dubbed them ELFs (Extremely Luminous FIR sources)  while others dubbed them ULIRGs (Ultra-Luminous IR Galaxies). Sadly, the acronym ULIRG stuck, and the world was deprived of papers discussing the relationship between ELFs and monsters.
 A debate raged whether such ULIRGs were powered by star formation or Active Galactic Nuclei (AGN).

ULIRGs are relatively rare at low z, but dominate cosmic star formation at high redshift \citep{wilman08}. Understanding their nature is therefore a crucial step to understanding the evolution of galaxies, but their high obscuration by dust makes it very difficult to study their innards, as optical and near-infrared observations often penetrate only the outer layer, and tell us little of their nucleus. 

In some cases, ULIRGs represent a transitional stage in which gas-rich spirals are merging to form a dusty quasar \citep{armus87, sanders88, spoon09}. In this scenario, this merger fuels the pre-existing quiescent black holes and triggers a powerful nuclear starburst, causing high IR luminosity, and generating strong starburst-driven winds which will eventually blow away the enshrouding dust and lay bare the quasar core, depleting the dust and gas to form an elliptical galaxy. 

The most luminous ULIRG discovered by IRAS was a source named
F00183-7111 (hereafter 00183). At a redshift of 0.3276 \citep{roy97}, it has a luminosity of 10$^{13}$\Lo, most of which is radiated at
far-infrared wavelengths. At the time of its discovery, it was the most luminous object then known
in the Universe.

\cite{spoon04} showed that it contains a cocoon of dense warm gas enclosing a very compact energy source, and \cite{roy97} found a strong radio-loud AGN at its centre with a unusually high  radio luminosity. As a result, 00183 is one of the  best-studied ULIRGs, with substantial amounts of time on Spitzer, Herschel, and ground-based telescopes allocated to its study.

\begin{figure}
\centering
\includegraphics[width=10cm]{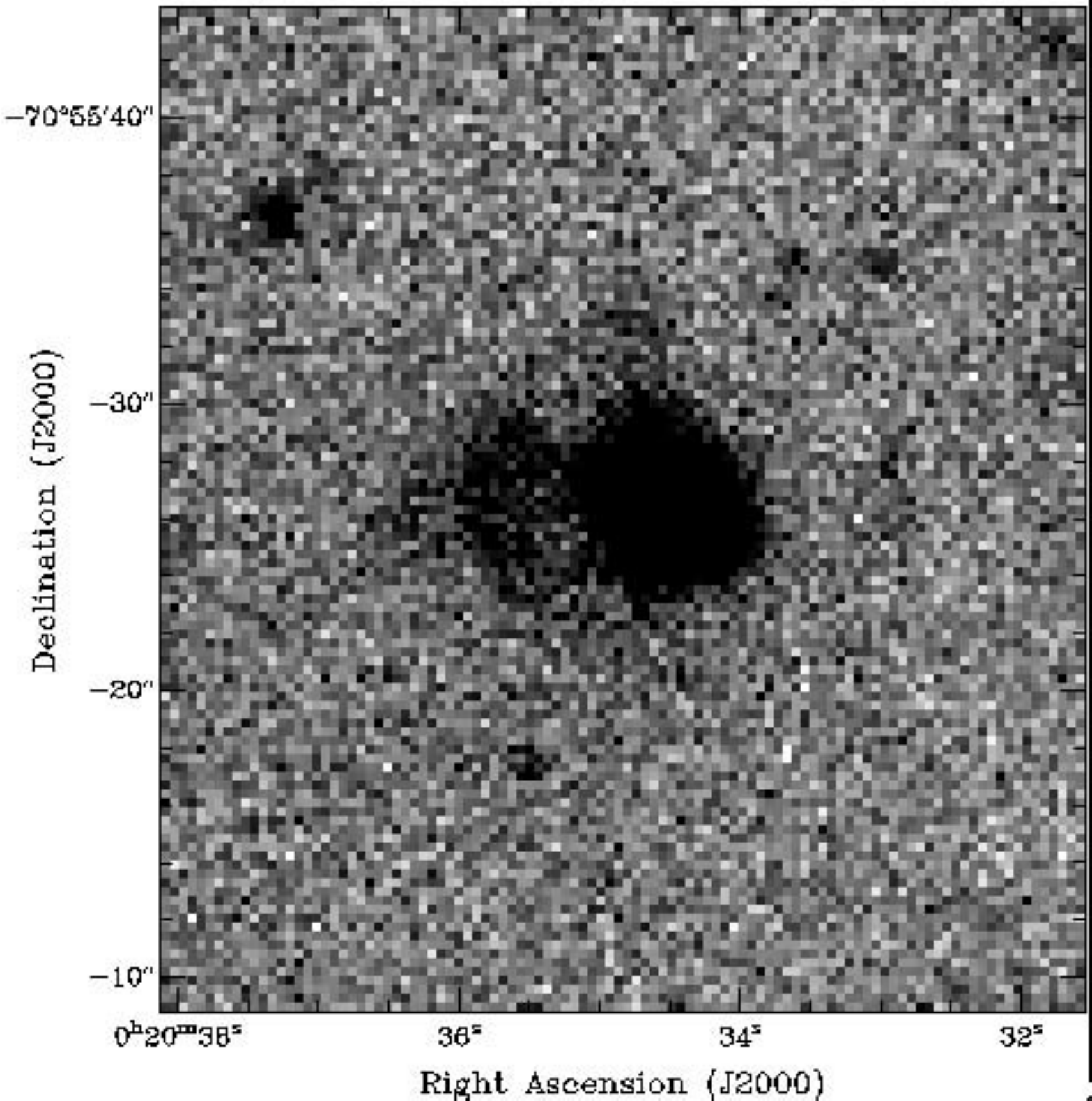}
\caption{Uncalibrated R+I band image of IRAS F00183-7111, taken with the 2dF focal plane imager on the Australian Astronomical Telescope (AAT) in June 2012. The entire disturbed galaxy has a total extent of about 15  arcsec.
} 
\label{fig1}
\end{figure}

Near-infrared imaging by \cite{rigopoulou99} revealed a disturbed morphology and a single nucleus, while observations by \cite{dannerbauer05} showed  extreme [FeII] emission. Long-slit spectroscopy \citep{drake04} showed bright, highly-disturbed, ionised gas extending over a distance of 12  arcsec.  In Fig. \ref{fig1} we present a recent image obtained on the AAT, showing what appear to be tidal tails, 
or remnants from a recent merger.  However, the strong dust extinction towards the nucleus, even at near-IR wavelengths, implies that these observations are telling us about the outer parts of the galaxy rather than the nucleus.

\begin{figure}
\centering

\includegraphics[width=10cm]{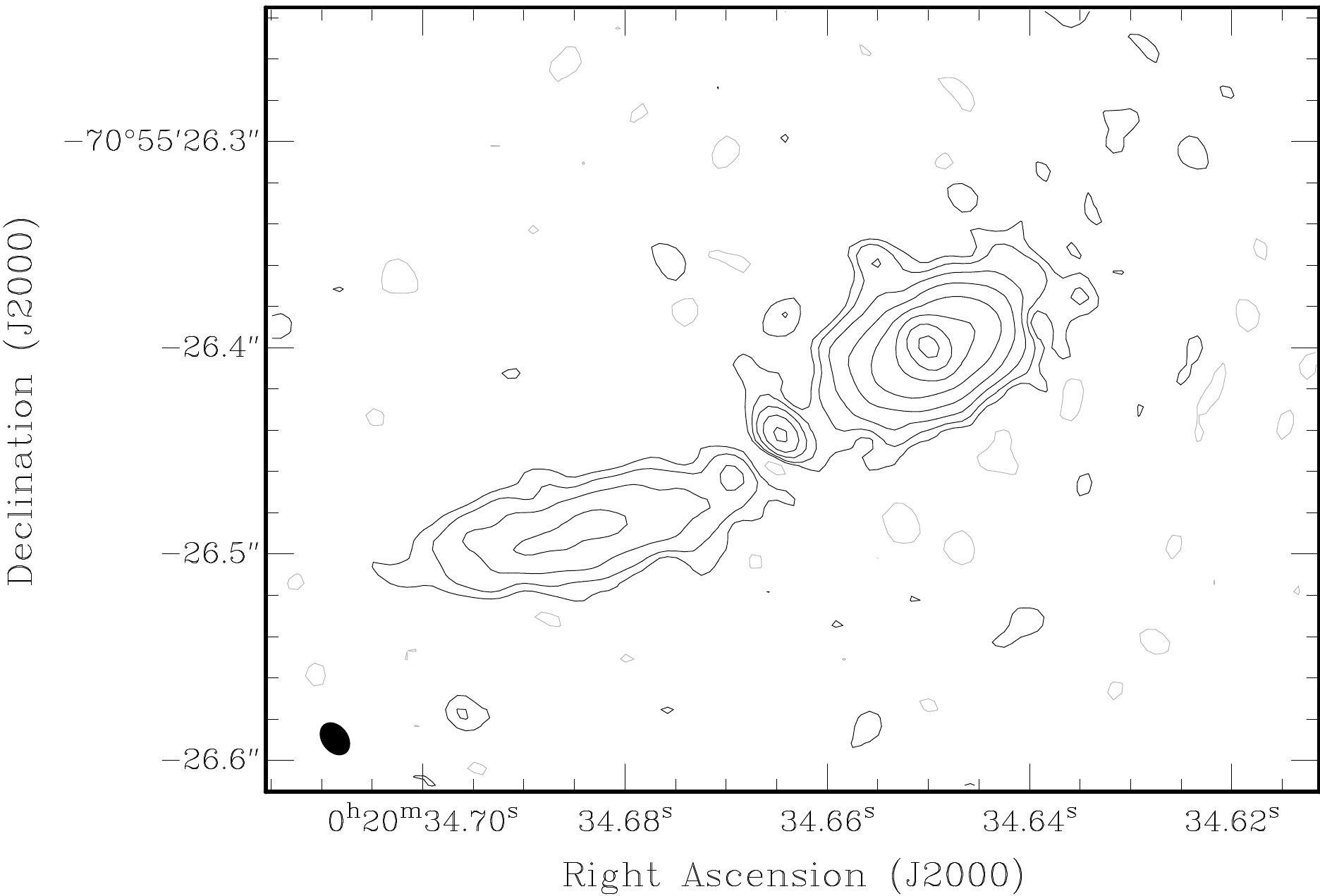} 
\caption{The VLBI image of F00183-7111, taken from Norris et al. (2012). The entire radio source is just over 1  kpc in extent, but is embedded in a dusty starburst some 20 kpc in extent.
} 
\label{fig2}
\end{figure}

\section{The AGN core}
Infrared and optical observations demonstrated unequivocally that 00183 is dominated at those wavelengths by an intense starburst. But \cite{roy97} showed that it is also a strong radio source, lying far above the radio-FIR correlation, suggesting that at radio wavelengths it is dominated by an AGN. High-resolution radio observations with the ATCA showed an unresolved radio core, whilst IR and optical observations showed a fuzzy blob with no hint of an AGN, although subsequent X-ray observations, and most recently mid-infrared observations \citep[e.g.][]{spoon09}, also supported the suggestion of an AGN. 

\cite{roy97} showed that the radio source at the centre of 00183 has a 
radio luminosity of
 $L_{6cm} = 3 \times 10^{25} W Hz^{-1}$, placing it in the regime of high luminosity (FRII-class) radio galaxies, a radio luminosity which is unusual in ULIRGs. Roy et al. suggested that the AGN is hidden at optical and near-infrared wavelengths by the dense dust galaxy surrounding it. The prevailing wisdom at the time was that radio-loud objects are almost always hosted by elliptical galaxies, so finding a radio-loud AGN embedded within a star-forming galaxy was a surprising result, although it is now well-accepted that ``composite'' galaxies showing both AGN and vigorous star-forming activity are widespread at high redshift.
 
 This radio AGN has recently been confirmed   by a VLBI image \citep{norris12a}, which shows a classical core-jet morphology (see Fig. 2). While its radio luminosity of $L_{20cm}$ of $\sim 6.10^{25} W~Hz^{-1}$ places it in the class of the most luminous (FRII) radio galaxies, most such galaxies have lobes spanning hundreds of  kpc, whereas the lobes of 00183 are only $\sim$1.5  kpc across. According to current models \citep[e.g. ][]{shabala11} such small jets have only just switched on and are boring their way through the dense dust and gas, injecting energy into it as they do so. 

The spectral energy distribution (SED) of the radio source also gives us a clue to its age, with a constant spectral index of -1.49 from 86 GHz down to a few GHz, and then turning over at around 1 GHz. Such a spectrum is characteristic of the sources known as CSS/GPS (compact steep-spectrum/gigahertz-peaked spectrum \citep{odea98, randall11}  which are widely thought to represent an early stage of  evolution of a classical FRII radio galaxy 
 
Thus, not only is this ULIRG radio loud, but the morphology and radio SED both  suggest that the AGN has just turned on and the tiny but powerful radio jets are still confined to the centre of the galaxy. We appear to have caught a ULIRG in the  brief period as a radio-loud AGN switches on in the centre and starts boring its way through the dense dust and gas.

\section{The Molecular Gas}
In 2007, we observed 00183 at 3mm with the Australia Telescope Compact Array (ATCA), and obtained a clear detection of the 3-mJy continuum from the nucleus, but only a marginal detection of a CO (1-0) line. Further observations of the CO gave confusing results, as the CO flux seemed to vary in a strange way as we changed the observing frequency. We suspected that we had detected a broad CO line, but decided not to publish, as the results were confusing. 

With the advent of the CABB upgrade of the ATCA correlator \citep{wilson, emonts11}, we obtained much better data which confirmed our detection. These new data \citep{norris12b}, shown in Fig 3, also showed why our earlier results had been so confusing, as the line was many times wider than the pre-CABB ATCA bandwidth. 
Although Fig. 3 clearly shows a detection, it is too noisy to tell us the width, central velocity, or integrated flux of the CO. A derived $H_{2}$ mass is very uncertain because it is unclear how much of the emission to the right of the peak in Fig. 3 is real, and whether the usual assumptions about $CO/H_{2}$ are valid in this unusual source, but is likely to exceed
  $M_{H2} =  10^{10}$ \Mo.

The peak of the CO in Fig. 3 is redshifted by about 500 km/s with respect to the centroid of most other lines. However Fig. 3 (and also our recently-acquired unpublished optical spectroscopy) shows that high velocities are widespread in this system, presumably partly as a result of  its recent violent history, partly because of the AGN, and partly because of starburst-driven winds, to the extent that it is difficult to define a ``systemic velocity''. The CO spectrum may also be affected by bulk motions of different components of this galaxy, or by absorption by foreground gas.  

Fig. 4 also shows that there may be a slight spatial offset between the CO and the continuum, suggesting that the AGN may not be at the centre of the molecular gas. This would not be surprising given the disruption of the host evident in Fig. 1.

\begin{figure}
\centering
\includegraphics[width=10cm]{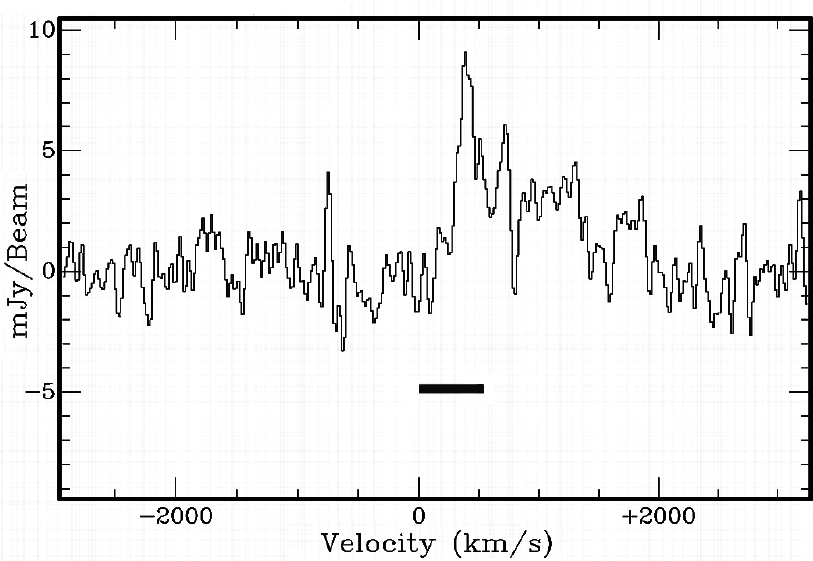} 
\caption{
Spectrum of CO (1-0) obtained with the ATCA (Norris et al. 2012b), showing an 8-$\sigma$ detection obtained near the rest frequency of the galaxy. However, the quality of the data, obtained soon after CABB commissioning, is insufficient to tell whether the apparent emission to the right of the peak is real or a baseline artefact. 
Velocity is relative to an assumed systemic redshift of z=0.3280, which is close to the centroid of the emission at most optical and IR wavelengths. A continuum signal of $\sim$ 3 mJy has been subtracted from this spectrum.
The bar below the spectrum shows the total bandwidth of the pre-CABB ATCA observations. 
} 
\label{fig3}
\end{figure}

\begin{figure}
\centering
\includegraphics[width=10cm]{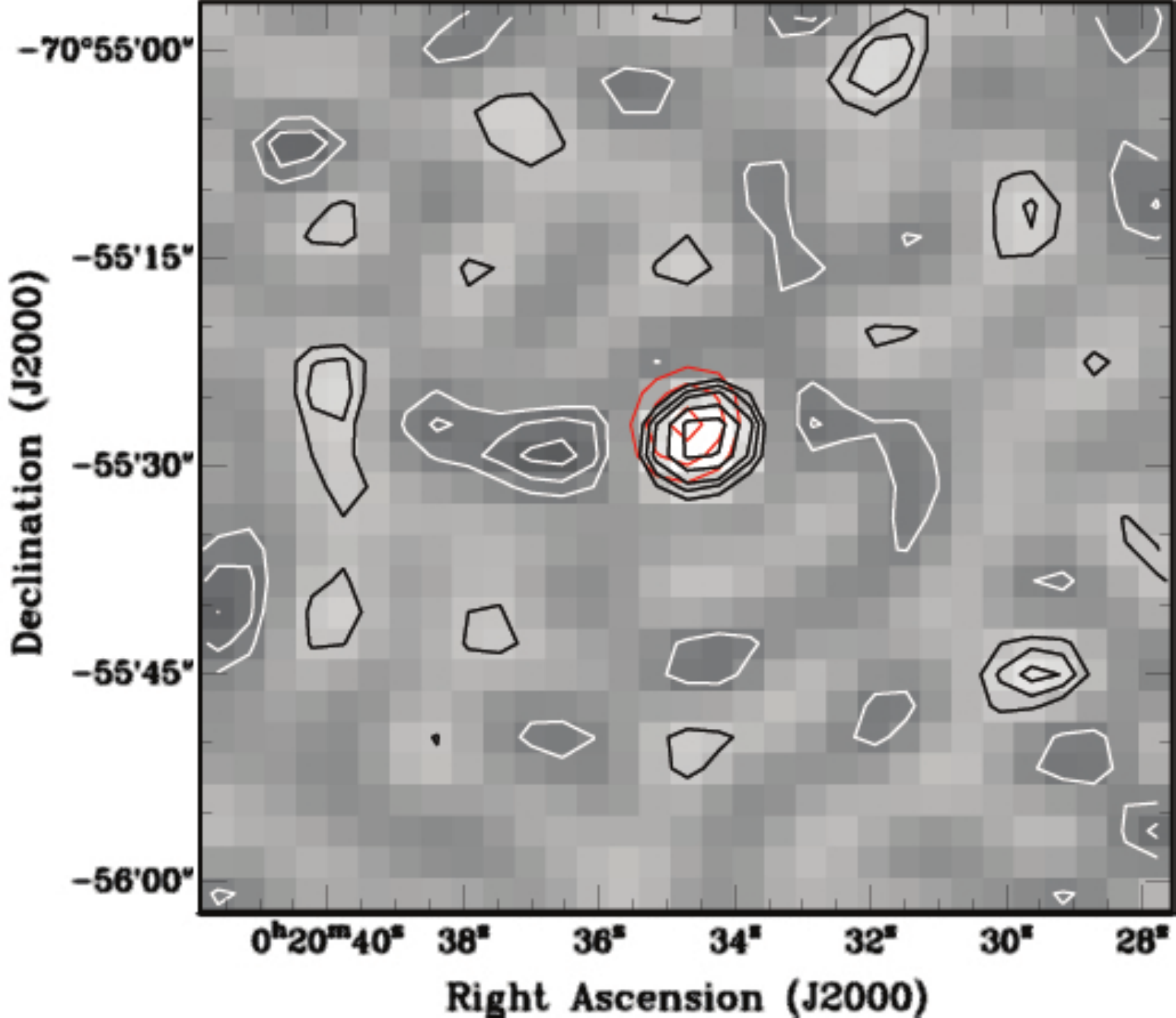} 
\caption{ Zero-moment image from the same data as used to generate Fig. 3. Black contours show the CO emission, and red contours show the continuum emission. 
} 
\label{fig4}
\end{figure}

\section{Discussion}

With hindsight, we realise that the standard model of AGN formation predicts sources such as 00183, caught at the earliest stages of AGN formation, and paint the following scenario. Two gas-rich spirals have merged, and the resulting heap of debris is undergoing an enormous starburst, seen at optical and IR wavelengths. At the centre of the merging spirals are one or more black holes from the nuclei of the merging galaxies, although models do not yet tell us whether these black holes have already merged or are still in the process of merging. Presumably the strong knot between the two jets is the black hole from the nucleus of one of the merging galaxies. It is tempting to speculate that the weaker knot at the base of the eastern jet may be the nucleus of the other merging galaxy, but confirmation of this must await a higher dynamic range VLBI image.

These black holes are being fed prodigious amounts of cold gas from the spirals (in "quasar accretion mode" in the terminology of \cite{croton06}), producing the high-luminosity jets  seen in the VLBI image. However, this process is invisible at optical and IR wavelengths, being hidden by the hundreds of magnitudes of extinction by the dust surrounding the nucleus.

This same dense dust and gas envelops the jets, preventing them from bursting out of the galaxy, and so the jets are gradually boring their way through the cold gas, heating and disrupting the material as they do so. Eventually they will burst out, heating the gas, and quenching the starburst. The source will then become a radio-loud quasar (or radio galaxy, depending on orientation), and will eventually settle down to radio-mode accretion (i.e. fuelled by hot gas).

In 00183 we are therefore probably witnessing a brief transition period between starburst-driven ULIRG and a quasar.  Although models are not yet sufficiently sophisticated to predict the length of this transition phase, the rarity of objects like 00183 attest to its brevity. Furthermore, 00183 appears to be a low-redshift analogue of the high-redshift radio galaxies, offering us the opportunity to study one of these objects with much higher resolution and sensitivity than is available at higher redshifts. So it's important to use every technique at our disposal to understand this object.

We have proposed ALMA observations, which will be about 15 times more sensitive, will cover a much wider frequency range, and will image the emission with a spatial resolution 4 times better than the ATCA. As a secondary goal, it will also produce a continuum image with about 20 times the sensitivity of the ATCA.

 Eventually we hope to use other molecules as well as CO, using the long baselines available in later ALMA phases to peel away the layers of obscuring gas and understand the kinematics of even the deepest layers surrounding the nucleus. Then, perhaps, we will start to understand the processes that turn merging spirals into quasars, and build the massive black holes at their centres.
 
Although 00183 is extreme, it is not unique, and similar compact radio sources  have been found buried within  lower-luminosity ULIRGs such as PKS1345+12 \citep{lister03} and PKS1549-79 \citep{holt06}. In the longer term, next generation radio surveys such as EMU \citep{norris11} will doubtless uncover many more examples of this stage of galaxy evolution. Given a sufficiently large sample, with accompanying multiwavelength data, it should be possible to sequence them to establish the early stages of radio-loud AGN and examine how their properties vary with the nature of the hosts and their environment.

\end{document}